\documentstyle[psfig]{mn}

\newif\ifAMStwofonts

\def\lapp{\ifmmode\stackrel{<}{_{\sim}}\else$\stackrel{<}{_{\sim}}$\fi}
\def\gapp{\ifmmode\stackrel{>}{_{\sim}}\else$\stackrel{>}{_{\sim}}$\fi}

\title[Two pulsars towards the Galactic Centre]
{Discovery of two pulsars towards the Galactic Centre}
\author[Johnston et al.]
{Simon Johnston$^1$, M. Kramer$^2$, D. R. Lorimer$^2$, A. G. Lyne$^2$,
\newauthor M. McLaughlin$^2$, B. Klein$^{3}$ and R. N. Manchester$^1$\\
$^1$Australia Telescope National Facility, CSIRO, P.O. Box 76, 
Epping, NSW 1710, Australia. \\
$^2$ University of Manchester, Jodrell Bank Observatory, Macclesfield, Cheshire, SK11 9DL.\\
$^3$ Max-Planck Institut f\"ur Radioastronomie, Auf dem H\"ugel 69,
53121 Bonn, Germany.\\
}
\date{\today}
\pagerange{\pageref{firstpage}--\pageref{lastpage}}
\pubyear{2006}
\begin{document}
\maketitle
\label{firstpage}

\begin{abstract}
We report the discovery of two highly dispersed pulsars in the direction
of the Galactic Centre made during a survey at 3.1~GHz with the 
Parkes radio telescope. Both PSRs~J1745$-$2912 and J1746$-$2856
have an angular separation from 
the Galactic Centre of less than 0.3\degr\ and dispersion 
measures in excess of 1100~cm$^{-3}$pc, placing them in the top 10 pulsars
when ranked on this value.
The frequency dependence of the scatter-broadening in PSR~J1746$-$2856
is much shallower than expected from simple theory.
We believe it likely that the pulsars are located between 150 and 500~pc
from the Galactic Centre on the near side, and are part of an excess
population of neutron stars associated with the Centre itself.
A second survey made at 8.4~GHz did not detect any pulsars. This
implies either that there are not many bright, long-period pulsars
at the Galactic Centre or that the scattering is more severe
at high frequencies than current models would suggest.
\end{abstract}


\section{Introduction}
Surveys for radio pulsars have been extremely successful over the
past decade \cite{lor05}.
All-sky surveys and deep surveys of the Galactic plane have
doubled the total number of known radio pulsars to above 1700.
At the same time,
targeted surveys have discovered very young pulsars
in supernova remnants, old millisecond pulsars in globular clusters
and a population in the Small and Large Magellanic Clouds.
In spite of these successes, no pulsars are known within $\sim$1\degr\
of the Galactic Centre (GC).

Following the Johnston et al. (1995)\nocite{jwv+95} targeted search of the 
GC at 1500~MHz (no detections), the most sensitive 
search at 1500 MHz is the Parkes multibeam survey which integrated for 
35~mins over a large fraction of the Galactic plane including the 
GC. That survey discovered PSR~J1747$-$2802, which has a period of 2.8~s,
the smallest angular separation from the
GC of all the known pulsars ($\sim$1\degr) and a high 
dispersion measure (DM) of 835~cm$^{-3}$pc \cite{mhl+02}.
At higher frequencies, the most sensitive survey was undertaken with
the Effelsberg telescope which
surveyed a small region (0.2 degrees or 32 pc radius) around the 
GC at 4850~MHz \cite{kkl+00,kle04}. In spite of a sensitivity 
below 100~$\mu$Jy and the high observing frequency
the survey failed to discover any new pulsars.

Sgr A* is at the heart of our Galaxy and is believed to be 
a black hole with a mass of $3\times 10^6 M_{\odot}$ \cite{egos02}
at a distance of $\sim$8~kpc.
A nearby stellar cluster appears to contain early-type stars with
masses of $10-20 M_{\odot}$. Two of the stars in this cluster 
are in highly eccentric orbits about the black hole with periods
of 15 and 30 years \nocite{gsh+05}. It is likely that shorter orbits
exist, however confusion in the infra-red has so far prevented their
detection. The presence of these high mass stars and
young supernova remnants is a good indicator
that active radio pulsars are also likely to exist close to the GC.
Pfahl \& Loeb (2004)\nocite{pl04} estimate that there are 1000 radio pulsars
within a few pc of the GC, a small fraction of which
are potentially detectable using current instruments provided that
scatter broadening does not render them invisible at any sensible
observing frequency (see also Cordes \& Lazio 1997\nocite{cl97}).
However, it is likely that the entire GC volume out to
$\sim$200~pc contains an overabundance of pulsars generally. Estimates
show that about 10\% of all the high mass stars (the pulsar progenitors)
in the Galaxy are contained within this volume \cite{frk+04}.

One can therefore
expect to detect a population of pulsars which would be extremely useful
in probing the GC and its conditions:
their number and age
distribution would probe the past star formation history \cite{har95};
their period derivatives can constrain the
gravitational potential in the GC;
pulsar timing would enable us to probe the
space-time around the super-massive black hole in the GC
due to a variety of relativistic effects \cite{wk99}.
The high stellar density of the GC
makes it, like the globular clusters, a possible site of a
millisecond pulsar orbiting a stellar-mass black hole though these
will be extremely difficult to detect.
Pulsar timing can also probe the
plasma density in the central regions through measurements of the
(variable) dispersion measure of a pulsar as it orbits the 
central black hole.

We therefore embarked on two surveys of the GC at frequencies
of 3.1 and 8.4~GHz, frequencies which bracket that of the Effelsberg
5~GHz survey. The lower frequency reduces the effects of scatter
broadening by a factor of $\sim$16
compared to surveys at 1.5~GHz, but at the same time the telescope
beam size is large enough to survey a significant area of sky in a 
reasonable amount of time. At the higher frequency,
the small beam restricts the sky coverage and the flux densities
of pulsars are smaller but the scatter broadening is lower by a factor
of more than 1000 compared to 1.4~GHz. Cordes \& Lazio (1997)\nocite{cl97}
identified frequencies near 8~GHz as the ideal for this type of search.

\section{Observations and Data Reduction}
The surveys were carried out using the Parkes radio telescope 
at centre frequencies of 3.1 and 8.4~GHz. At the lower frequency
the total bandwidth was 576~MHz in
each of two polarizations which was subdivided into 192 channels
each of width 3~MHz.
At 8.4~GHz, 288 channels were employed for a total bandwith of 864~MHz.
The outputs from the channels were 1 bit digitised and sampled every
250~$\mu$s and subsequently written to disk for off-line processing.
Figure~1 shows the surveys areas superposed on a
continuum image of the GC region.

Observations at 3.1~GHz were carried out from 2005 July 19 to 22.
A total of 32 pointings were made, each with an integration time of 70~mins. 
At this frequency, the half-power width of the telescope beam is 7 arcmin.
The survey therefore covered 0.34 square degrees on the sky or a box
approximately 90~pc across at the distance of the GC.
On cold sky, the system equivalent flux density was $\sim$45~Jy, as measured 
through observations of the calibrator source Hydra~A.
However, conditions near the GC contribute substantially
to this value. From the maps of Reich et al. (1984)\nocite{rfh+84},
we estimate a contribution of $\sim$25~Jy in the outer regions of
the survey, $\sim$65~Jy in the inner regions and up to $\sim$550~Jy
at the GC itself.  For pulsars with
a duty cycle of 10\%, the detection threshold (10-$\sigma$) is
then $\sim$120~$\mu$Jy (outer regions), $\sim$190~$\mu$Jy (inner regions)
and $\sim$1~mJy at the GC.

Observations at 8.4~GHz were carried out from 2005 September 13 to 16.
The survey involved 31 pointings, each observed for 70~mins.
The half-power width of the telescope beam is 2.4 arcmin and the
survey covered 0.04 square degrees on the sky.
The system equivalent flux density on cold sky was 48~Jy. Additions to
this from emission at the GC were estimated from 
the maps of Seiradakis et al. (1989)\nocite{srw+89} to be $\sim$4~Jy
in the outer regions of the survey, $\sim$10~Jy in the inner regions
and $\sim$100~Jy at the GC.
The 10-$\sigma$ detection threshold is then $\sim$70~$\mu$Jy (outer regions),
$\sim$80~$\mu$Jy (inner regions) and $\sim$200~$\mu$Jy at the GC.

Data reduction was carried out using the
{\sc sigproc}\footnote{{http://sigproc.sourceforge.net}}
software package. An initial
pass through the data involved resampling to 1~ms and applying 415
(at 3.1~GHz) or 62 (at 8.4~GHz) trial
dispersion delays to the data for a range of dispersion measures (DMs)
up to 10000~cm$^{-3}$pc. A 2$^{22}$ point fast Fourier transform was 
then done on the dedispersed time series and the resultant power spectrum 
searched for significant spikes. Harmonic summing in 4 stages up to a 
factor of 16 was carried out and the most significant signals written to disk.
These were then time-folded to produce a candidate pulse profiles for
subsequent visual inspection.

A search for isolated dispersed bursts of emission with
signal-to-noise ratios above a 5-$\sigma$ threshold was also performed
\cite{mll+06}. Time series were smoothed with boxcars of various widths
to increase our sensitivity to broadened pulses. No sources of bursts
were found, though the high frequency of these surveys and hence
the relatively low dispersion delay, makes
distinguishing radio frequency interference from signals of astrophysical 
origin difficult.

\section{Two new pulsars}
Two periodicities, near 945 and 187~ms, were stand-out candidates from the
data reduction of the 3.1~GHz data, with signal to noise ratios of
41.5 and 9.4. Confirmation of the first pulsar, PSR~J1746$-$2856, came
from analysis of archival data from the Parkes multi-beam survey at 1.4~GHz.
The 945~ms pulsar is highly scattered at that frequency but is clearly
detected with a signal to noise ratio of 16.
A re-detection at 3.1~GHz was made at Parkes on 2005 August 26.
The second pulsar, PSR~J1745$-$2912, was not seen in the archival 
1.4~GHz data but was confirmed at Parkes at 3.1~GHz on 2005 August 27.
Successful detection of both pulsars at 5~GHz was also made
using the Effelsberg telescope in early 2005 September, and more accurate
positions were obtained by performing a grid search around the discovery
locations.
The location and pulse profiles for both pulsars are shown in Figure~1.
\begin{table}
\caption{Parameters for PSRs J1745$-$2912 and J1746$-$2856}
\begin{tabular}{lll}
\hline & \vspace{-3mm} \\
 & J1745$-$2912 & J1746$-$2856 \\ 
R.A. (J2000) & 17$^{\rm h}$ 45$^{\rm m}$ 50(10) & 17$^{\rm h}$ 46$^{\rm m}$ 49\fs5(5) \\
Dec (J2000)  & $-$29$^\circ$ 12\arcmin (2) & $-$28$^\circ$ 56\arcmin 31\farcs0(2) \\
Gal. longitude (deg) & 359.79 & 0.12 \\
Gal. latitude (deg) & $-$0.18 & $-$0.21 \\
P (ms) & 187.3794(2) & 945.224316(3) \\
$\dot{P}$ ($\times10^{-15}$) & & 12.5(3) \\
DM (cm$^{-3}$pc) & 1130(20) & 1168(7) \\
Epoch & 53609.30 & 53704.50932 \\
S$_{3.1}$ (mJy) & 0.12 & 0.76 \\
Age (Myr) & & 1.2 \\
B ($\times10^{12}$~G) & & 3.5 \\
\hline & \vspace{-3mm} \\
\end{tabular}
\end{table}

The pulsars have very large DMs; only 14 pulsars were previously known
with DMs in excess of 1000~cm$^{-3}$pc and of these only PSR B1758$-$23 is
within 15 degrees of the GC.
Scattering times were estimated using a technique described in
L\"ohmer et al. (2001)\nocite{lkm+01}. We assumed that there is no
significant scattering at 4.8~GHz and used the pulse profiles obtained
at Effelsberg as an estimate of the true pulse profile.
We then convolved this profile with a truncated exponential with one
free parameter (the scattering time) and compared it to the observations
at lower frequencies using least-squares fitting.
For PSR~J1745$-$2912 we obtain a scattering time of $25\pm3$~ms at 3.1~GHz
and for PSR~J1746$-$2856 the scattering time is $15\pm2$~ms at 3.1~GHz and
$170\pm15$~ms at 1.4~GHz. This implies the power-law index of the
scattering as a function of frequency is $-3.0\pm0.3$, significantly 
smaller in magnitude
than the expected value of $-4.4$, but in line with the results obtained
for other high DM pulsars \cite{lkm+01}. Extrapolating to 1~GHz with
a $-3$ index, gives values of 750 and 450~ms respectively.

Following the confirmation of the pulsars, subsequent timing observations
of PSR~J1746$-$2856 were carried out using the Lovell telescope
at the Jodrell Bank Observatory at a frequency of 1.4~GHz with additional
data at 3.1~GHz from the Parkes telescope.
A total of 49 observations since mid-2005 have allowed
allowed us to determine a timing solution for this pulsar.
As yet, a timing solution has not
been obtained for PSR~J1745$-$2912 due mainly to its lack of 
detectability at 1.4~GHz. Parameters for both pulsars are listed in Table~1.
\begin{figure}
\psfig{figure=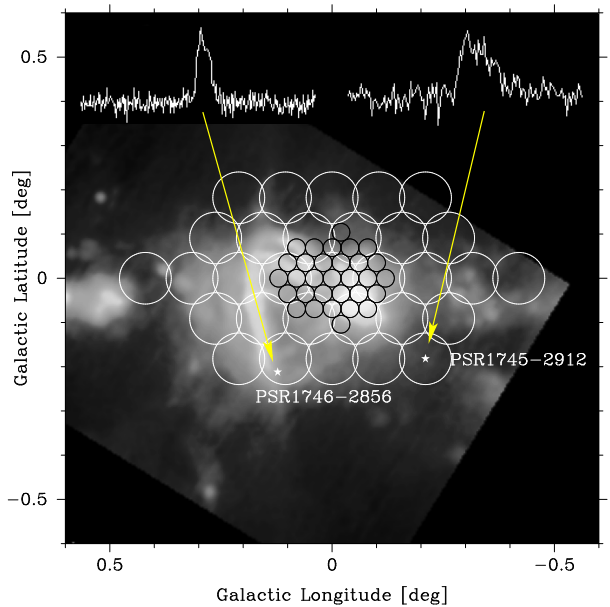,angle=0,width=8cm}
\caption{Image of the Galactic Centre region showing the surveyed regions
and the profiles of the newly discovered pulsars. The background image
shows continuum emission at a frequency of 10.55~GHz from the
Effelsberg survey of Seiradakis et al. (1989).
The larger circles denote individual survey pointings at 3.1~GHz
with the smaller circles representing the pointings for 
the 8.4~GHz survey. The two stars mark the positions of the discovered
pulsars, with their respective pulse profiles shown at the top.
Note that both pulsars lie outside the region surveyed both at 8.4~GHz and 
at 5~GHz with Effelsberg.}
\end{figure}

\section{Expected DM and scattering at the Galactic Centre}
Sgr~A* itself is known to be extremely scatter broadened and this is
also true of maser and extragalactic sources at small angular separations
from the GC. Although the exact nature and location of the screen
which causes the scattering is not known, there are good reasons to
believe that is it very close to the GC.
Cordes \& Lazio (1997)\nocite{cl97} and Lazio \& Cordes
(1998)\nocite{lc98} have shown that
any pulsars located beyond the GC scattering screen
should be extremely scatter broadened, rendering them undetectable
in conventional pulse searches at frequencies below a few GHz.
They estimate that the scattering screen lies at a distance of $\sim$130~pc
from the GC, encloses the GC and is likely to be patchy.
The estimates of the scattering times and DMs of pulsars at the 
GC are $\sim$400~s (at 1~GHz) and $\sim$2000~cm$^{-3}$~pc.

The two pulsars are both at an angular separation of 0.3\degr\ from
the GC, corresponding to $\sim$40~pc lateral displacement.
If they had a similar radial displacement from the GC, they would be 
well inside the putative scattering screen. Their measured scatter broadening
values of $\sim$500~ms would therefore be about two orders 
of magnitude lower than expected.
This discrepancy is made even worse if the pulsars are
located behind the GC. Furthermore, the measured DMs
are a factor of 2 less than expected for the GC.
We can also examine the scatter broadening of OH maser spots in the
GC. The two OH masers nearest PSR~J1745$-$2912 are
OH359.762+0.120 and OH359.880$-$0.087 with angular sizes of 
1700 and 2800~mas (scaled to 1~GHz) respectively \cite{vfcd92,fdcv94}.
We can convert our measured time scatter broadening values of $\sim$750~ms
at 1~GHz to an angular broadening and obtain a value of 700~mas assuming
a distance of 8.5~kpc and a scattering screen located close to the pulsar.
Again this likely indicates the pulsars are in front of the GC.

Both the Taylor \& Cordes (1993)\nocite{tc93} and
Cordes \& Lazio (2002)\nocite{cl02} models of the electron density
consist of an outer, thick disk component, an inner, annular 
component, and spiral arms. Although the models differ significantly in
some parts of the Galaxy, they give very similar results in the inner
$\sim$1~kpc, largely due to the lack of constraints there.
If the pulsars are indeed in front of the scattering screen but within
the inner few hundred pc, the expected DM would then be only
$\sim$650~cm$^{-3}$~pc in both of the models listed above.
Scattering is expected to be $\sim$50~ms (at 1~GHz).
The measured values of the DM and the scattering time are
significantly higher than the models would suggest, perhaps indicating
that a filled centre model, rather than an annular one may be
more appropriate.

In any case, it seems most likely that the pulsars would have to
be located in front of the scattering screen and at least 150~pc from
the GC.  This still leaves the question as to their
place of birth; in the sections below we consider possible origins for the
birth location of these pulsars.

\section{Location and origin of the newly discovered pulsars}
\subsection{Field pulsars: Formation at $D>$1~kpc from the GC}
There is a possibility that these pulsars are merely field pulsars
on the near side of the GC but not associated with
any (excess) neutron star population there. However,
the 3.1~GHz survey would not expect 
to find any pulsars from the normal field
population for two reasons. First, the survey area is tiny;
the Parkes multibeam survey detected pulsars at a rate of $\sim$1 per
square degree in the inner Galaxy whereas our survey 
covered only 0.34 square degrees
with a roughly similar effective sensitivity (taking into account the
survey parameters, the background temperature and the spectral index
of pulsars).
Secondly, there is good evidence that the number of field pulsars per unit
area actually {\it decreases} close to
the GC \cite{joh94,yk04,lor04} making detections even
more unlikely in this volume.

We have performed a simulation of the normal Galactic population of
pulsars using the $z$-height and radial distributions of
Johnston (1994) and the space
velocity distribution of Lyne \& Lorimer (1994) with a random orientation
of the birth kick.  We allow a random (flat)
distribution of ages up to a maximum of 10$^7$~yr and let the pulsars
evolve in the Galactic potential.  In this simulation
we find only 0.1\% of the pulsars are located in the region
covered by the 3.1~GHz survey.
In contrast, if we simulate a population of GC 
neutron stars with a Gaussian radial distribution with $\sigma=70$~pc,
we find that some 10\% of the sample remains inside the survey region.
There are estimates that about 10\% of all high mass star formation
takes place in the inner few hundred pc of the Galaxy. If this is the
case then, in the region covered by the 3.1~GHz survey, one would
expect to have 10 times more pulsars with origins from the 
GC population compared to those pulsars which have migrated
inwards from the outer Galaxy.

The similarities in the parameters of these two pulsars are striking.
Their DMs and scattering times are similar, and they have a similar
angular separation from the GC. This seems unlikely to be
random chance. Taking all this information together, it
therefore seems highly probable
that both the newly discovered pulsars are part of an `extra' population 
directly closely linked to conditions at the GC.

\subsection{Formation within 500~pc of the GC}
The pulsars could have been
born in the stellar cluster which surrounds Sgr A* and occupies
about only 0.02~pc \cite{pl04}. In this case, their orbital speed around
the GC would be $\sim$1000~km~s$^{-1}$
and the birth kick would be unlikely to perturb them significantly from
their orbit. This possibility therefore seems unlikely.

The pulsars could originate from a progenitor population within
$\sim$1~pc of the GC. At this
distance, the gravitational effect of the black hole is small and 
the birth process could kick the pulsars out to their current position.
A pulsar with a velocity of 100~km~s$^{-1}$ travels 100~pc in 1~Myr,
the characteristic age of PSR~J1746$-$2856,
and no especially large kick velocity needs to be invoked.
In this picture, the pulsars' proper motion 
would be in a direction away from the GC.

The pulsars could also originate
from a distance of $\sim$200~pc from the GC inside of which there is 
known to be significant high mass star formation \cite{frk+04}. The velocity
of the pulsars would be moderate so as to ensure their retention
in the GC region. Long term timing of the two pulsars may help to
distinguish between these two cases.

\section{No detections at 8.4~GHz}
What are the implications of failing to detect any pulsars at 8.4~GHz?
Recall that the area surveyed is only a $\pm$15~pc box around the
GC. At the GC distance, the survey region is inside the scattering screen
proposed by Lazio \& Cordes (1998) and any putative pulsars will therefore
suffer from broadening. However, assuming the screen is 130~pc
from the GC, the scattering time at 8.4~GHz is then expected to
be only $\sim$80~ms.

Ignoring this effect for now,
the sensitivity of the survey is such that a pulsar
with a luminosity (at 8.4~GHz) greater than 7~mJy~kpc$^2$ would have
been detected. Extrapolating down to 1.4~GHz, assuming a spectral index
of $-$1.6, gives a luminosity limit of 125~mJy~kpc$^2$. Of the
1160 known pulsars with a flux density measurement at 1.4~GHz, 133
of them (11\%) have luminosities in excess of this. Given that there
are $\sim$10$^5$ active pulsars beaming in our direction and that
the known population is not complete even at this luminosity level,
at least 0.1\% of all pulsars should have luminosities
in excess of this value. Therefore, either less than 1000 pulsars
beaming in our direction are in the GC region or, as seems more
likely, the scattering is severe even at this high frequency.
Note that in the inner 1~pc, the luminosity limit is a factor of 3 
higher because of the high background temperatures there (see Section 2).
This increases the upper limit to the number of detectable pulsars
by a similar amount. This is then broadly in line with modelling
of the GC population of neutron stars \cite{cl97,pl04}.

\section{Summary}
We have discovered two pulsars within 0.3\degr\ of the GC 
during a targeted survey at 3.1~GHz. Both pulsars have very high 
DMs and scatter broadening times.
It seems unlikely that the survey has penetrated
through the scattering screen which surrounds the GC at a distance
of $\sim$150~pc. However, the pulsars are likely to have originated
in an excess population associated with the GC and be
located within a few hundred pc of it. Both the DM and scattering are
then higher than expected in the current electron density models
perhaps favouring a filled centre model rather than the existing
annular one. No detections were made at 8.4~GHz with the implication being
that either there are less than 3000 pulsars beaming in our direction
in the inner pc of the Galaxy or that the scattering is 
more severe than previously thought.

\section*{Acknowledgments}
The Australia Telescope is funded by the Commonwealth of 
Australia for operation as a National Facility managed by the CSIRO.
DRL is a University Research Fellow funded by the Royal Society.
We thank John Reynolds for his assistance with the instrumental setup
and Wolfgang Reich for providing the 11cm continuum map of the GC.
We acknowledge use of the ATNF pulsar catalogue located at
http://www.atnf.csiro.au/research/pulsar/psrcat.

\bibliography{modrefs,psrrefs,crossrefs}
\bibliographystyle{mn}
\label{lastpage}
\end{document}